\documentclass[reprint,showpacs,preprintnumbers,tightenlines,aps,prx]{revtex4-1}
\usepackage{graphicx}
\usepackage{multirow}
\usepackage{latexsym,amssymb}
\usepackage{amsmath,bbold}
\usepackage{paralist}
\usepackage{braket}

\newcommand{\del}[2]{\partial_{#2}{ #1}}

\newcommand{\dprime}[1]{#1^{\prime\prime}}

\begin{document}

\title{Boundary twists, instabilities, and creation of skyrmions and antiskyrmions}
\author{Aldo Raeliarijaona}
\affiliation{Department of Physics and Astronomy and Nebraska Center for Materials and Nanoscience, University of Nebraska, Lincoln, Nebraska 68588, USA}
\author{Rabindra Nepal}
\affiliation{Department of Physics and Astronomy and Nebraska Center for Materials and Nanoscience, University of Nebraska, Lincoln, Nebraska 68588, USA}
\author{Alexey A. Kovalev}
\affiliation{Department of Physics and Astronomy and Nebraska Center for Materials and Nanoscience, University of Nebraska, Lincoln, Nebraska 68588, USA}
\date{\today}

\begin{abstract}
We formulate and study the general boundary conditions dictating the magnetization profile in the vicinity of an interface between magnets with dissimilar properties. Boundary twists in the vicinity of an edge due to Dzyaloshinskii-Moriya interactions have been first discussed in [Wilson \textit{et al.}, Phys. Rev. B 88, 214420 (2013)] and in [Rohart and Thiaville, Phys. Rev. B 88, 184422 (2013)]. We show that in general case the boundary conditions lead to the magnetization profile corresponding to the N\'eel, Bloch, or intermediate twist. We explore how such twists can be utilized for creation of skyrmions and antiskyrmions, e.g., in a view of magnetic memory applications.
To this end, we study various scenarios how skyrmions and antiskyrmions can be created from interface magnetization twists due to local instabilities. We also show that a judicious choice of Dzyaloshinskii-Moriya tensor (hence a carefully designed material) can lead to local instabilities generating certain types of skyrmions or antiskyrmions. The local instabilities are shown to appear in solutions of the Bogoliubov-de-Gennes equations describing ellipticity of magnon modes bound to interfaces.
In one considered scenario, a skyrmion-antiskyrmion pair can be created due to instabilities at an interface between materials with properly engineered Dzyaloshinskii-Moriya interactions. We use micromagnetics simulations to confirm our analytical predictions.
\end{abstract}

\maketitle

\section{Introduction} 
Since its first theoretical proposal by Bogdanov and R\"{o}{\ss}ler \citep{Bogdanov-Robler,Robler-Bogdanov-Pfeiderer}, and first experimental discovery in cubic B20 compound MnSi \citep{Mbauer}, skyrmions have attracted flurries of research interest from the community. The size of skyrmions can be as small as 20 nm \citep{Fert-Review} which allows to densely store information.
Typically, skyrmions are stabilized by the Dzyaloshinskii-Moriya interaction (DMI) \cite{DzyaloshinskyJPCS1958,MoriyaPR1960} which is present in the absence of the center of inversion as a result of spin-orbit interactions.
The fidelity of skyrmion memory is guaranteed by the stability of the skyrmionic state, which owes this property to the topological nature of skyrmions \citep{Nagaosa-Review}. The skyrmions can be controlled electrically \citep{Hsu,White}, mechanically via acoustic waves \citep{Rabindra}, via uniaxial stress \citep{Nii}, or thermally \citep{Romming,Kong,Kovalev:PRB2014,Mochizuki}.
Furthermore, multiferroicity in such materials as GaV$_{4}$S$_{8}$ \citep{Ruff} enables the low-energy encoding and decoding of information via skyrmions. Skyrmion-based logic gates proposed recently \citep{XZhang} rely on the ability to change a skyrmion with positive charge into a skyrmion with a negative charge.
Alternatively, helicity of skyrmion can be changed by passing it through region with position dependent DMI \citep{Diaz-Troncoso} which can be achieved by varying the level of doping \cite{Shibata.Yu.ea:NN2013,Siegfried.Altynbaev.ea:PRB2015}.
Recently, it has been demonstrated that the strength of DMI can be controlled by electric field in complex oxide films \cite{Ohuchi.Matsuno.eaNC2018}.

In thin magnetic films, skyrmions are typically stabilized by interfacial DMI \cite{Crepieux.LacroixJoMaMM1998,Bogdanov.HubertJoMaMM1994,Bogdanov.RoeslerPRL2001,EzawaPRB2011,Kiselev.Bogdanov.eaJoPDAP2011}.
Recent theoretical studies have established the relation between the asymmetry of interfacial DMI and the existence of skyrmions and antiskyrmions \citep{Utkan,Hoffmann}. In particular, it has been demonstrated that skyrmions can be stabilized in chiral magnets with Rashba-like DMI and antiskyrmions can be stabilized in chiral magnets with Dresselhaus-like DMI.
The contemporary progress in experimental capabilities, such as molecular beam epitaxy \citep{Ahmed_MBE}, or pulsed-laser deposition \citep{Schlenhoff_PLD} demonstrate the ability to control material growth on a layer by layer basis. In principle, such control in growth can enable the generation of layered chiral magnets with tailored properties, such as crystal symmetry, leading to different types of DMI, and hence providing a way to enable formation of skyrmions or antiskyrmions. 
Recent experiments also show that DMI can be engineered via the capping heavy metal \cite{Balk.Kim.eaPrl2017,Wells.Shepley.eaPRB2017}. Alternatively, antiskyrmions can be also stabilized by dipole-dipole interactions \cite{Koshibae.Nagaosa:NC2016} in the presence of the anisotropy  created by ion irradiation in Co/Pt multilayers \citep{ZPP}. Electrical control of the magnetocrystalline anisotropy can also be used for skyrmion stabilization \citep{Maruyama}.

Edge magnetization twists due to DMI have been discussed by Wilson \textit{et al.} \cite{Wilson.Karhu.eaPRB2013} and by Rohart and Thiaville \cite{Rohart-Thiaville}. 
In a semi-infinite slab of chiral magnet, the edge can assist in the creation of non-trivial spin structures such as chiral spin states, or skyrmions \citep{Rohart-Thiaville,Muller-Rosch-Garst,Garanin.Capic.ea2018}. The proper boundary conditions play an important role in describing local instabilities \citep{Rohart-Thiaville,Muller-Rosch-Garst,Hals.Everschor-Sitte:PRL2017}. 
The boundary conditions have only been studied for the interfacial or bulk DMI. Here, we study such boundary conditions for a general DMI tensor and show that such generalized boundary can lead to formation of magnetization twists.
As has been demonstrated, a skyrmion-antiskyrmion pair can be created by current pulse in a conventional chiral magnet where antiskyrmion eventually disappears \citep{Stier}. In the present study, we investigate the behavior of magnetization near an interface separating two magnetic regions with dissimilar DMI, or a boundary with anisotropic DMI and demonstrate generation of stable skyrmions and antiskyrmions. We first formulate the general boundary conditions that must be obeyed by the magnetization at an interface between materials with different properties. 
To identify local instabilities leading to the creation of skyrmions and antiskyrmions, we study the spin wave excitations at the interface or edge by diagonalizing the Bogoliubov-de-Gennes Hamiltonian within the continuum model. 
We support our analytical investigation with micromagnetics simulations that confirm creation of skyrmions and antiskyrmions at the interface or edge of the sample.

The paper is organized as follows. In Sec.~\ref{sec:Method}, we review the edge boundary conditions and derive the general interface boundary conditions. In Sec.~\ref{sec:EqSpinD}, we use the boundary conditions to obtain the magnetization twists at interfaces and edges. In particular, we identify the N\'eel- and Bloch-type twists. In Sec.~\ref{sec:Magnon}, we study magnon modes localized on edges and interfaces by solving the Bogoliubov-de Gennes equations. We show that such magnon modes can lead to local instabilities and formation of non-collinear states. In Sec.~\ref{sec:microM}, we use micromagnetics simulations to demonstrate the creation of skyrmions and antiskyrmions from local instabilities. Our concluding remarks are presented in Sec.~\ref{sec:Summary}.

\section{Method}\label{sec:Method}
\subsection{Free energy functional}
In our study, we assume that the tensor describing DMI can be position dependent, e.g., as a result of doping \cite{Shibata.Yu.ea:NN2013,Siegfried.Altynbaev.ea:PRB2015} and/or variations in the capping layer \cite{Balk.Kim.eaPrl2017,Wells.Shepley.eaPRB2017}. For a 2D system well below the Curie temperature, we write the magnetic free energy, $F=\int d^d \boldsymbol r {\mathcal F}$, and its density
\begin{equation}\label{FreeE}
{\mathcal F}=(J_{\alpha\beta}/2)\partial_{\alpha}\boldsymbol m \cdot \partial_{\beta}\boldsymbol m -K m_{z}^{2}-H m_{z}+ D_{i\alpha k}   m_i \partial_{\alpha}  m_k,
\end{equation}
where we assume summation over repeated indices, $\alpha$ and $i(k)$, and $\boldsymbol m$ is a unit vector along the magnetization vector. The first term in Eq.~(\ref{FreeE}) describes exchange with the exchange stiffness $J_{\alpha\beta}$ which is usually assumed to be isotropic $J_{\alpha\beta}=J\delta_{\alpha\beta}$, the second term describes uniaxial anisotropy with the strength $K$, the third term describes the Zeeman energy due to the external magnetic field $H_e$, $H\equiv\mu_{0}H_e M$, and the last term corresponds to DMI described by a general tensor $D_{ijk}$. In what follows, we will often split the DMI tensor into symmetric and antisymmetric parts, $D_{ijk}=D^S_{ijk}+D^A_{ijk}$ where $D^S_{ijk}=(D_{ijk}+D_{kji})/2$ and $D^A_{ijk}=(D_{ijk}-D_{kji})/2$. Note that the rank-2 tensor $\mathcal{D}_{jm}=-\epsilon_{mik} D^A_{ijk}$, where $\epsilon_{mik}$ is the Levi-Civita symbol, is sufficient to describe the behavior of the system in the bulk \cite{Hals.Everschor-Sitte:PRL2017}. 

We consider an interface between two regions and assume that the exchange stiffness and DMI tensor can vary across this interface on a lengthscale that is smaller than the typical size of the magnetic texture. Under the assumtion of slowly varying magnetic texture defined by the weak spin-orbit interaction, the presence of interactions limited to interface should not affect the behavior of magnetization. Within our continuum approach, this can be seen by adding the interactions arising at interface on interatomic scales to the Free energy, i.e., $K^s m_i^2 \delta(x)$, $J^s \left(\partial_{\alpha}\boldsymbol m\right)^{2} \delta(x)$, and $D^s_{i\alpha k}   m_i \partial_{\alpha}  m_k \delta(x)$ where without loss of generality we assume that the interface is at $x=0$ and $\delta(x)$ is a delta function. Under the assumption of weak spin-orbit coupling and after using dimensionless units from relation $\delta(x)=(D/J)\delta( x D/J)$ it follows that all additional terms will be higher order in the strength of spin-orbit interaction \cite{Meynell.Wilson.ea:PRB2014,Muller-Rosch-Garst}. 

In our discussion, the magnetization dynamics will be described by the Landau-Lifshitz-Gilbert (LLG) equation:\begin{align}\label{LLG}
s(1-\alpha\boldsymbol m \times) \dot{\boldsymbol m} - \boldsymbol m \times \delta_{\boldsymbol m} F = 0,
\end{align}
where $F$ is the total free energy, $s=M_s / \gamma$ is the spin angular momentum density, and $\gamma$ is (minus) the gyromagnetic ratio ($\gamma>0$ for electrons). To derive Eq.~(\ref{LLG}) one could employ the variational principle in which the Gilbert damping corresponds to the Rayleigh dissipation function \cite{Gilbert:IToM2004}. 

\subsection{Edge boundary condition}
We first consider an edge of a magnetic layer with a general DMI. The boundary condition  corresponding to the Neumann boundary \cite{Rohart-Thiaville,Muller-Rosch-Garst} follows from a variational principle applied to the free energy  
$F=\int_{-\infty}^{\infty} d^{d-1} \boldsymbol r \int_0^{\infty} d x {\mathcal F}$ where a constraint $\boldsymbol m \cdot \boldsymbol m=1$ has to be imposed \cite{Hals.Everschor-Sitte:PRL2017}. Integration by parts results in the following general boundary condition \cite{Hals.Everschor-Sitte:PRL2017}:
\begin{equation}\label{BC1}
 n_i J_{ij} \del{\boldsymbol m}{j}+\boldsymbol m \times(\boldsymbol \Gamma_D \times \boldsymbol m)=0,
\end{equation}
where $\boldsymbol n$ is the normal pointing outside of the region and $(\Gamma_{D})_k=m_i n_j D^A_{ijk}$.  Below, we use this boundary condition in order to show a possibility to create skyrmions and antiskyrmions at an edge of a magnetic layer with anisotropic DMI.

\subsection{Interface boundary condition}
We also consider an interface between two regions where the first region is described by the exchange stiffness $J^L_{ij}$ and the DMI tensor $D^L_{ijk}$ and the second region is described by the exchange stiffness $J^R_{ij}$ and the DMI tensor $D^R_{ijk}$. As mentioned before, we assume that this change happens on a lengthscale that is smaller than the typical magnetic texture size. To obtain the boundary conditions, we split the free energy integration into two parts $F=\int_{-\infty}^{\infty} d^{d-1} \boldsymbol r \int_0^{\infty} d x {\mathcal F}+\int_{-\infty}^{\infty} d^{d-1} \boldsymbol r \int_{-\infty}^{0} d x {\mathcal F}$ and apply the variational principle to each of the two terms separately. The boundary terms result in a boundary condition:
\begin{equation}\label{BC2}
 n^L_i J^L_{ij} \del{\boldsymbol m}{j}+n^R_i J^R_{ij} \del{\boldsymbol m}{j}+\boldsymbol m \times\left[(\boldsymbol \Gamma_D^L+\boldsymbol \Gamma_D^R) \times \boldsymbol m\right]=0,
\end{equation}
where $\boldsymbol n^L$ and $\Gamma_D^L=m_i n^L_j D^A_{ijk}$ correspond to the left region, and $\boldsymbol n^R$ and $\Gamma_D^R=m_i n^R_j D^A_{ijk}$ correspond to the right region.
The boundary conditions in Eqs.~(\ref{BC1}) and (\ref{BC2}) will be employed in what follows to describe boundary twists at interfaces.

\subsection{Boundary-induced anisotropy}
In Eqs.~(\ref{BC1}) and (\ref{BC2}) we only included the antisymmetric part $D^A_{ijk}$ of DMI tensor. In the formal derivation, the symmetric part also enters the boundary conditions \cite{Hals.Everschor-Sitte:PRL2017}. Such terms lead to the Free energy contribution:
\begin{equation}\label{E_sym1}
\displaystyle
F^{S}=\int d^{d-1} \boldsymbol r n_j D^{S}_{i j k}  m_i  m_k \bigg|_{b1}^{b2},
\end{equation}
which corresponds to the boundary anisotropy and is typically second order in the strength of spin-orbit interaction. For smooth magnetic textures such contributions should lead to higher order corrections as has been discussed earlier.

\section{Boundary magnetization twists}\label{sec:EqSpinD}
\begin{figure}[!ht]
\centering
\includegraphics[scale=0.6]{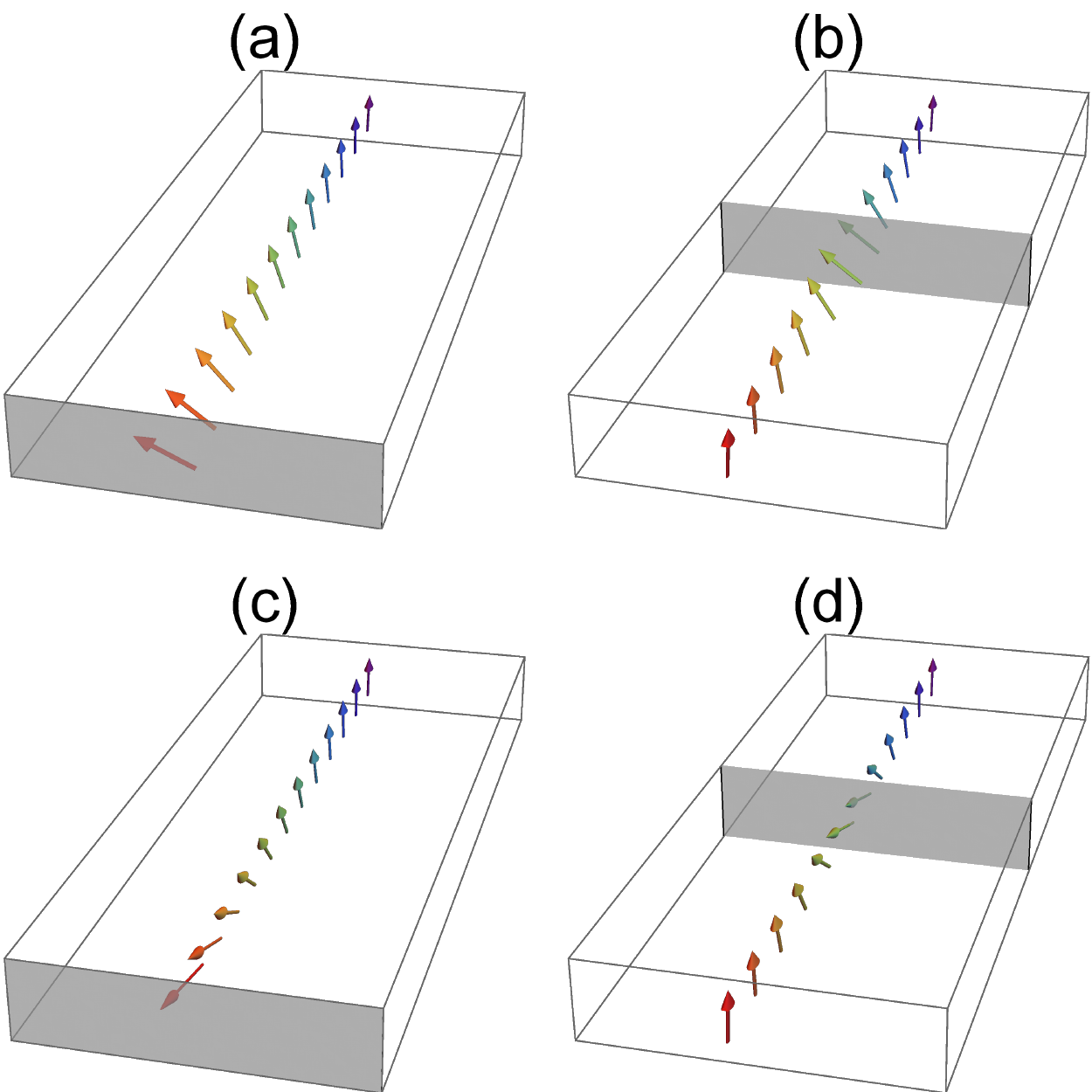}
\caption{(Color online) (a) A Bloch-type twist induced at an edge of a chiral magnet with DMI. (b) A Bloch-type twist induced at an interface between two chiral magnets by discontinuity in DMI. (c) A N\'eel-type twist induced at an edge of a chiral magnet with DMI. (d) A N\'eel-type twist induced at an interface between two chiral magnets by discontinuity in DMI. The gray region indicates where the edge or interface is located.}
\label{fig:cartoon}
\end{figure}

In this section, we study the static magnetization profile induced by DMI, generalizing results of Refs.~\cite{Muller-Rosch-Garst,Hals.Everschor-Sitte:PRL2017} to anisotropic DMI.  
The non-zero elements of DMI tensor are determined by relations:
\begin{equation} \label{eq:cryst}
{\cal D}_{ij}=(\det \boldsymbol R^{(\alpha)}) R^{(\alpha)}_{il} R^{(\alpha)}_{jm} {\cal D}_{lm},
\end{equation}
where $\boldsymbol R^{(\alpha)}$ are generators of the point group corresponding to the crystallographic symmetry, $\alpha=1,2,\ldots$, and the summation over repeated indices $l$, and $m$ is assumed. In analogy with magnetic domain walls, we identify the Bloch- and N\'eel-type twists as shown in Fig.~\ref{fig:cartoon}.

Before we proceed any further let us bring some clarification onto our notation of the DMI. We denote by $D_{ijk}$ the rank-3 DMI tensor as shown in the above equation; however as is customary in the literature \cite{Utkan}, the DMI rank-2 tensor is written as $\hat{\mathcal{D}}$. Often due to the symmetry constraints, only few entries in tensor $\hat{\mathcal{D}}$ are nonvanishing. 

\subsection{Analytical results}\label{subsec:analyt}

In the absence of the symmetric component, i.e. $D^S_{ijk}=0$, we can represent DMI by a rank-2 tensor. To uncover the most important physics, we limit our consideration to the following general DMI tensor:
\begin{equation}\label{ADMI_x}
\hat{{\mathcal D}}=
\begin{pmatrix}
D_{1} &  D_{3} & 0 \\
D_{4} &  D_{2}  & 0 \\
0 &  0 & D_{5} 
\end{pmatrix},
\end{equation}
where some elements are taken to be zero, e.g., due to the symmetry constraints \cite{Utkan}. For example,  for $C_{2v}$ symmetry we also have $D_1=D_2=D_5=0$.

We assume that in the bulk the magnetization is uniformly polarized along the z-axis, $\boldsymbol m=\hat{z}$. Near the boundary or interface at $x=0$, the magnetization will tilt due to the boundary conditions (\ref{BC1}) and (\ref{BC2}). 
Because of the translational invariance of the system along the y-direction the magnetization does not depend on $y$. We can thus consider the following ansatz for the magnetization ${\boldsymbol m}$:
\begin{equation}\label{n_ansatz}
\boldsymbol m^{T}=\left(\sin(\theta(x))\cos(\phi),\sin(\theta(x))\sin(\phi),\cos(\theta(x))\right),
\end{equation}
where $T$ indicates a transposed vector, and $\theta(x)$ and $\phi$ correspond to parametrization of $\boldsymbol m$ in terms of spherical coordinates. Note that $\phi=0$ corresponds to the N\'eel-type twist and $\phi=\pi/2$ corresponds to the Bloch-type twist. 

The static spin density is obtained from the variation of the Free energy with respect to $\boldsymbol m$, which results in the following equation written in dimensionless units for the function $\theta(x)$:
\begin{gather}\label{DSG}
\dprime{\theta}-\frac{\kappa}{2}\sin[2\theta(x)]-h \sin[\theta(x)]=0,
\end{gather}
where $h=H/(JQ^{2})$ is the external magnetic field, $\kappa=2K/(JQ^{2})$ is the uniaxial anisotropy, and $Q=\sqrt{D_1^2+D_4^2}/J$. In these dimensionless units $x$ is redefined as $Qx$. The value of $\phi$ is found from the boundary conditions.

The solutions to the double Sine-Gordon Eq.~(\ref{DSG}) take the form:
\begin{equation}\label{ThetaInt}
\theta(x)=\mp\pi\pm 2\tan^{-1}\left({\frac{\sqrt{h} \sinh\left\lbrace\sqrt{h+\kappa} (x- x_{0})\right\rbrace}{\sqrt{h+\kappa}}} \right),
\end{equation}
where the kink center $x_{0}$ is determined from the boundary conditions.
Note that for the case of an interface between two regions with different DMI we will need two solutions on both sides of the interface. It is Eq.~(\ref{ThetaInt}) that describes the tilting of the magnetization close to a boundary.
\begin{figure}[!ht]
\centering
\includegraphics[scale=0.9]{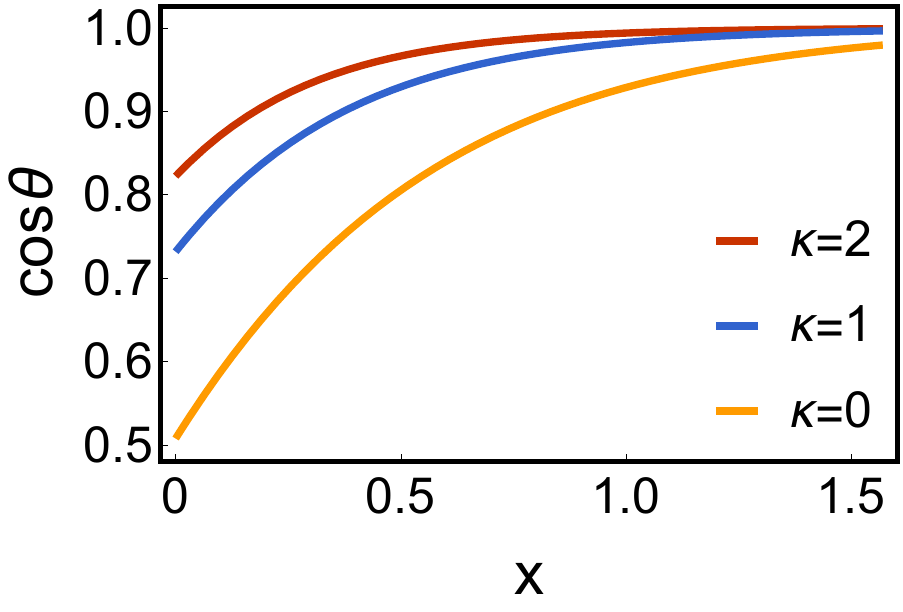}
\caption{(Color online) Numerically obtained magnetization profile (z-component of the unit magnetization vector) as a function of the dimensionless distance from the edge of a chiral magnet with $C_{\infty}$ symmetry. We use dimensionless parameters, $h=1$, $D_1=0$, and $D_4/(QJ)=-1$. The plot corresponds to the N\'eel-type twist with $\phi=\pi$. Dimensionless units are used.}
\label{fig:edge}
\end{figure}
\begin{figure}[!ht]
\centering
\includegraphics[scale=0.9]{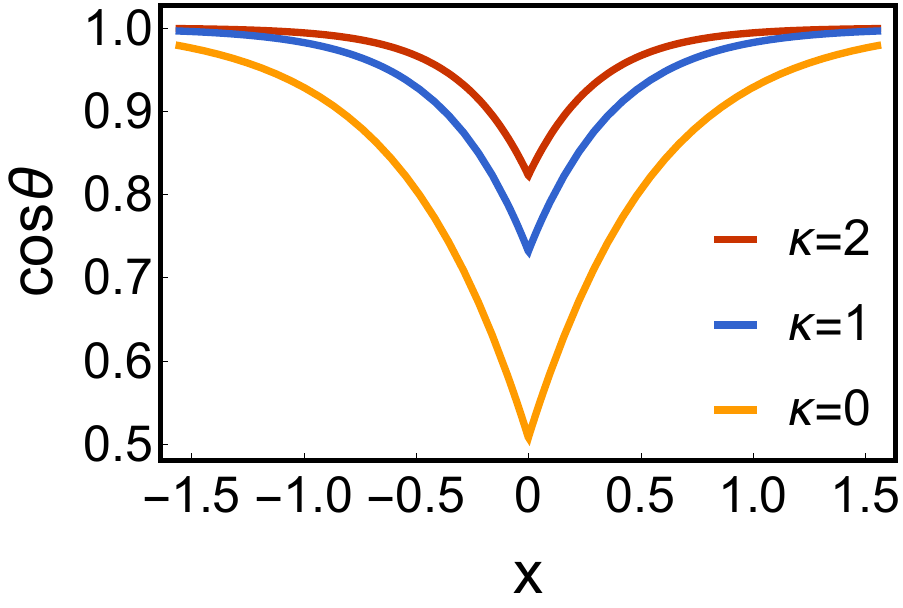}
\caption{(Color online) Numerically obtained magnetization profile (z-component of the unit magnetization vector) as a function of the dimensionless distance from the interface between chiral magnets with $C_{\infty}$ symmetry. We use dimensionless parameters, $h^L=h^R=1$, $\kappa^L=\kappa^R=\kappa$, $D_1^L/(QJ)=-D_1^R/(QJ)=-1$, and $D_4^L=D_4^R=0$. The plot corresponds to the Bloch twist with $\phi=\pi/2$. Dimensionless units are used.}
\label{fig:int}
\end{figure}
\subsubsection{Semi-infinite slab geometry}
We consider a semi-infinite chiral magnet in a region $x>0$ where the boundary is at $x=0$ along the y-axis.
Using Eq.~(\ref{n_ansatz}), the boundary condition (\ref{BC1}) leads to the following equations:
\begin{align}\label{Edgebc}
 \theta^{\prime}\rvert_{0^+}&=1, \\
 \sin(\phi)&=\frac{D_1}{\sqrt{D_1^2+D_4^2}},\\
 \cos(\phi)&=\frac{D_4}{\sqrt{D_1^2+D_4^2}}.
\end{align} 
The boundary condition (\ref{Edgebc}) then results in solution (\ref{ThetaInt}) where the position of the kink $x_{0}$ is given by equation:
\begin{equation}\label{InitCond}
x_{0}=-\frac{\cosh^{-1}\left( \frac{(h+\kappa)+\sqrt{(h+\kappa)^{2}-\kappa }} {\sqrt{h}}\right)}{\sqrt{(h+\kappa)}},
\end{equation}

We also obtain numerical results and compare them to analytical results. We use the shooting method in order to numerically find the stationary solution of LLG equation in spherical coordinates satisfying the boundary values. We observe a perfect agreement with analytical results as can be seen in Fig.~\ref{fig:edge}. Note that the angle $\phi$ is also extracted from the numerical procedure.

\subsubsection{Interface separating regions with different DMI}
\begin{figure}[!ht]
\centering
\includegraphics[scale=0.9]{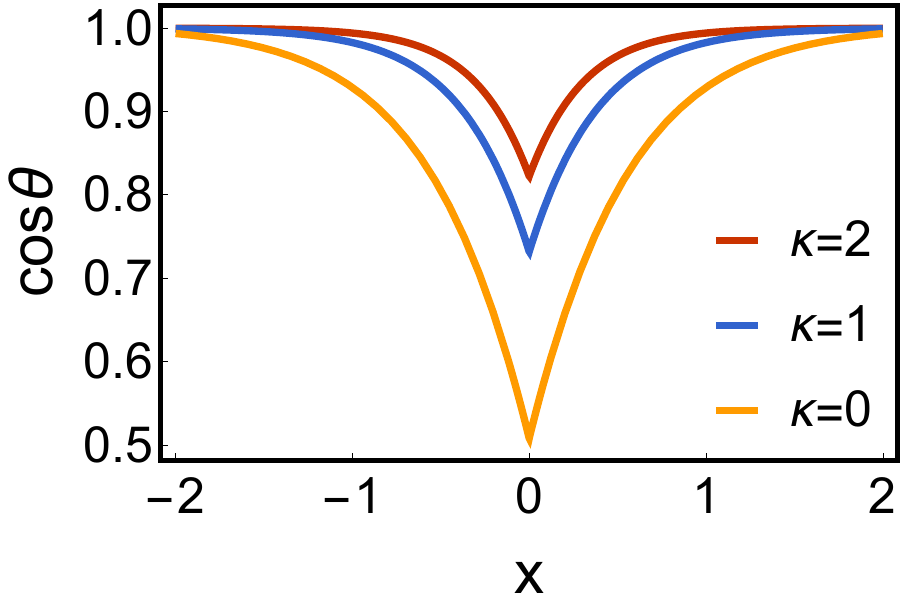}
\caption{(Color online) Numerically obtained magnetization profile (z-component of the unit magnetization vector) as a function of the dimensionless distance from the edge or interface of a chiral magnet with $C_{2v}$ symmetry. The interface corresponds to a chiral magnet with dimensionless parameters $h=1$ and $D_4/(QJ)=-1$. The numerically obtained angle $\phi$ corresponds to the N\'eel-type twist. Dimensionless units are used.}
\label{fig:C2v}
\end{figure}

We now consider an interface at $x=0$ separating two different regions. Because of the translational invariance of the system along the y-direction we again employ ansatz (\ref{n_ansatz}).  We use dimensionless units on each side where $Q^L=\sqrt{(D_1^{L}-D_1^{R})^2+(D_4^{L}-D_4^{R})^2}/2J^L$ and $Q^R=\sqrt{(D_1^{L}-D_1^{R})^2+(D_4^{L}-D_4^{R})^2}/2J^R$ define dimensionless coordinates $Q^L x$ and $Q^R x$ on each side. In general, parameters describing each region are given by $\kappa^L$, $\kappa^R$, $h^L$, and $h^R$ with Eq.~(\ref{DSG}) describing the magnetization profile on both sides. In addition to Eq.~(\ref{DSG}), each region also satisfies the boundary condition (\ref{BC2}) which after application of ansatz (\ref{n_ansatz}) leads to equations:
\begin{align}\label{Intbc}
 \theta^{\prime}\rvert_{0^+}&=2+\theta^{\prime}\rvert_{0^-}, \\
 \sin(\phi)&=\frac{D_1^R-D_1^L}{\sqrt{(D_1^{L}-D_1^{R})^2+(D_4^{L}-D_4^{R})^2}},\\
 \cos(\phi)&=\frac{D_4^R-D_4^L}{\sqrt{(D_1^{L}-D_1^{R})^2+(D_4^{L}-D_4^{R})^2}}.
\end{align}
From the boundary condition (\ref{Intbc}) we can recover the positions of the kinks for the left and right solutions (\ref{ThetaInt}), i.e. $x_0^L$ and $x_0^R$. The general analytical expressions for $x_0^L$ and $x_0^R$ are complicated and it is more practical to calculate them numerically. Nevertheless, for the case when only DMI varies across the interface, i.e. $h^L=h^R$ and $\kappa^L=\kappa^R$, we recover the analytical solution:
\begin{equation}\label{InitCond1}
x_{0}^{R/L}=\mp\frac{\cosh^{-1}\left( \frac{(h+\kappa)+\sqrt{(h+\kappa)^{2}-\kappa }} {\sqrt{h}}\right)}{\sqrt{(h+\kappa)}},
\end{equation}
Finally, we obtain numerical results and compare them to analytical results. We use the shooting method in order to numerically find the stationary solution of LLG equation in spherical coordinates satisfying the boundary values. We observe a perfect agreement with analytical results as can be seen in Figs.~\ref{fig:int} and \ref{fig:C2v}. Note that the angle $\phi$ is also extracted from the numerical procedure and it corresponds to the N\'eel- and Bloch-type twists, respectively.

\section{Skyrmions and antiskyrmions from boundary instabilities}\label{sec:Magnon}
In this section, we examine the spin-wave fluctuations around the equilibrium magnetization following the approach used in Refs.~\citep{Schuette.GarstPRB2014,Muller-Rosch-Garst} and identify edge and interface instabilities that can lead to formation of skyrmions and antiskyrmions. We focus on chiral magnets with $C_{2v}$ symmetries since such magnets can host both skyrmions and antiskyrmions \citep{Utkan,Hoffmann}. 
\begin{figure}[!ht]
\centering
\includegraphics[scale=0.9]{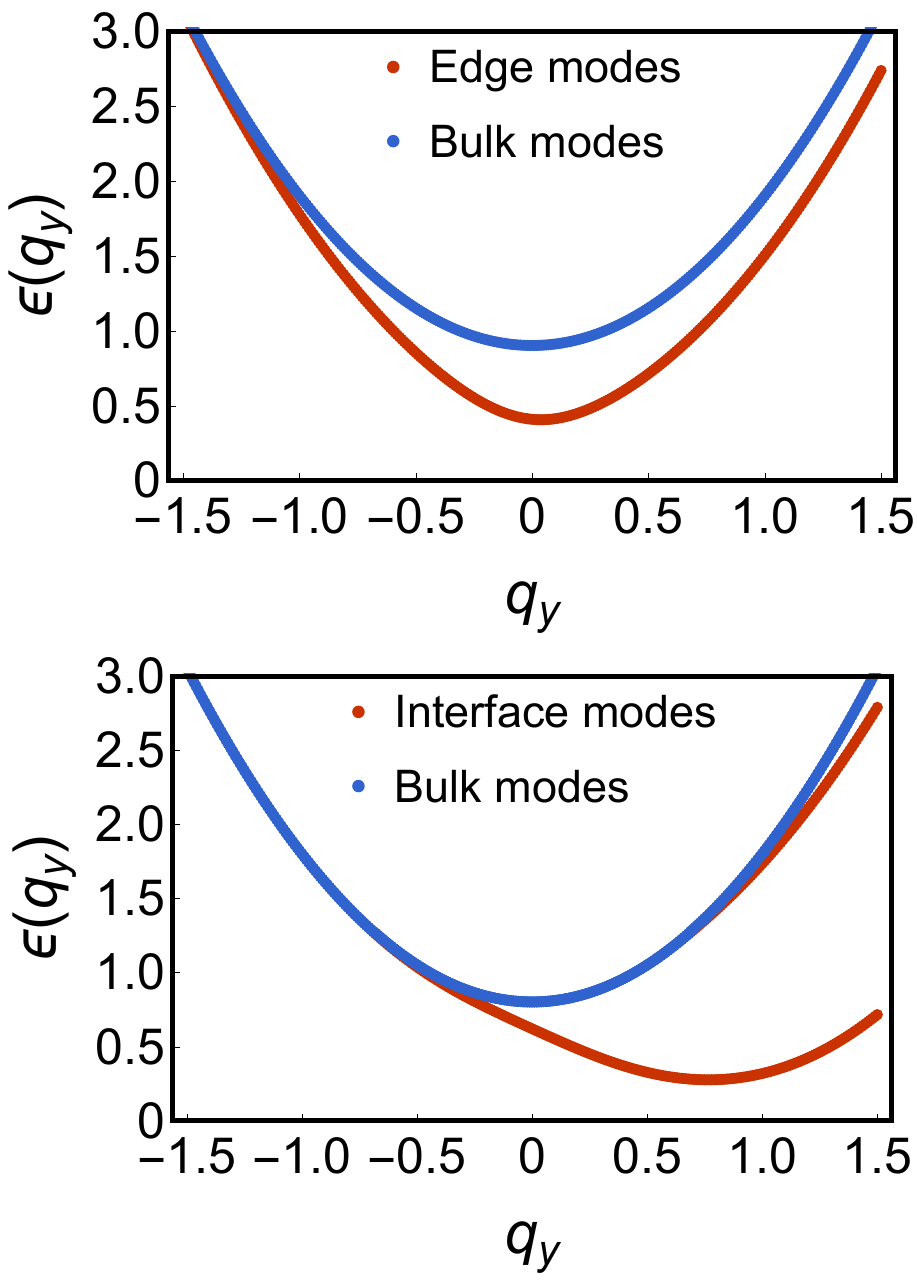}
\caption{(Color online) Upper plot: Dispersion plot for the edge and bulk magnon modes for $D_3/(QJ)=1$, $D_4/(QJ)=0.1$, $\kappa=0.8$, and $h=0.1$. Lower plot: Dispersion plot for the interface and bulk magnon modes for $D_3^L/(QJ)=1$, $D_4^L/(QJ)=-1$, $D_3^R/(QJ)=1$, $D_4^R/(QJ)=1$, $\kappa^L=\kappa^R=0.5$, and $h^L=h^R=0.3$. Dimensionless units are used.}
\label{Fig:Bands}
\end{figure}
\subsection{Instabilities at edges and interfaces}
For $C_{2v}$ symmetry with one of the mirror planes along the boundary, we always obtain the N\'eel-type twist. It is then convenient to introduce the orthogonal frame: 
\begin{gather}
\hat{\bf e}^{T}_{1}=(0,1, 0),\\
\hat{\bf e}^{T}_{2}=(-\cos[\theta(x)],0,\sin[\theta(x)]),\\
\hat{\bf e}^{T}_{3}=(\sin[\theta(x)],0,\cos[\theta(x)]),
\end{gather}
where $\theta(x)$ is the polar angle describing the twist.
We describe the spin-wave fluctuation of the equilibrium magnetization by a complex number $\psi(x,y,t)$, with $|\psi(x,y,t)|<<1$, where  the magnetization vector ${\boldsymbol m}$ can be parametrized as:
\begin{equation}
\hat{\boldsymbol m}=\hat{\bf e}_{3}\sqrt{1-2\mid \psi(x,y,t) \mid^{2}}+\hat{e}_{+}\psi(x,y,t)+ \hat{e}_{-}\psi^{*}(x,y,t).
\end{equation} 
Here $\psi^{*}(x,y,t)$ is the complex conjugate of $\psi(x,y,t)$ and $\hat{e}_{\pm}=\hat{e}_{1}\pm i \hat{e}_{2}$. Due to translational invariance along the y-direction we can use the Fourier transformed  spinor:
\begin{equation}
\psi(x,q_{y},t)=\int{dx e^{-iq_{y}y} \psi(x,y,t)},
\end{equation}
which reduces the problem to one dimension.
The eigenvalue equation is obtained after expanding the LLG equation (\ref{LLG}) to linear order in the fluctuation $\psi (x,y,t)$. The eigenvalue equation $H_{BdG}\Psi =\epsilon\tau^{z}\Psi$ with $\Psi=\left( \psi(x,q_{y},t),\psi^{*}(x,q_{y},t)\right)^{T}$ can be further simplified with the help of identities on the solutions of the double Sine-Gordon equation (\ref{DSG}). The resulting Bogoliubov-de-Gennes Hamiltonian \cite{Garcia-Sanchez.Borys.eaPRB2014,Muller-Rosch-Garst} can be decomposed as $H_{BdG}=H_{0}+V(x,q_y)$, where the so-called bulk contribution  $H_{0}$ reads:
\begin{equation}\label{BdGH0}
H_{0}=-\partial^{2}_{x} +q^2_{y} + (h+\kappa),
\end{equation}
and the potential $V(x,q_{y})$ depends on the particular form of the boundary or interface. We use the parametrization in Eq.~(\ref{ADMI_x})  ($D_1=D_2=D_5=0$) and results from the previous section for $C_{2v}$ case to obtain the general expression for the magnon potential $V^{L/R}(x,q_{y})$:
\begin{widetext}
\begin{gather}
\label{BdGVani}
V^{L/R}(x, q_{y})=\mathbb{1}(-{\theta^{\prime}}^2-\kappa^{L/R} \sin^{2}(\theta)- D_{4}^{L/R} \theta^{\prime} )+2 D_{3}^{L/R}\tau^{z}q_{y}\sin(\theta)+\tau^{x}\left(-\frac{{\theta^{\prime}}^{2}}{2}- D_{4}^{L/R} \theta^{\prime}+\frac{\kappa^{L/R}}{2}\sin^{2}(\theta)\right),
\end{gather}
\end{widetext}
where the indices $L$ and $R$ describe the potential to the left and to the right from the interface. The potential can be reduced to the one considered in Ref.~\citep{Muller-Rosch-Garst} by setting $D_3=-D_4=1$. 

For the interface, the eigenvalue problem is solved using the finite element method with the requirement that solutions decay as $x \rightarrow \pm\infty$. For the edge, an additional boundary condition $\del{\psi(x,q_y,t)}{x}=0$ follows directly from the general boundary conditions formulated earlier. Note that the normalization condition becomes:
\begin{equation}
\displaystyle
\int_{0}^{\infty}{\Psi^{\dagger}(x,y,t)\tau^{z}\Psi(x,y,t) dx}=1,
\end{equation}
where $\Psi^{\dagger}$ is the adjoint of $\Psi$ and $\tau^{z}$ is the third Pauli matrix. We use this eigenvector to track the magnon eigenenergy as a function of $q_y$.

In Fig.~\ref{Fig:Bands}, we plot the dispersion of magnon modes. The edge and interface bound modes have lower energy compared to the bulk modes with the bulk gap given by $\Delta_{b}=h+\kappa$. We observe that the energy of bound modes can becomes zero for some particular values of the magnetic field and anisotropy, at which point an instability develops. In our micromagnetic simulations, we confirm that this instability can lead to creation of skyrmions and antiskyrmions. Note that the upper plot in Fig.~\ref{Fig:Bands} describes an edge of a chiral magnet with highly anisotropic DMI. In principle, such anisotropic DMI is more favorable to the formation of chiral solitons. Nevertheless, in our micromagnetic simulations we observe that in some cases the soliton breaks into a skyrmion-antiskyrmion pair after the magnetic field is increased. The lower plot in Fig.~\ref{Fig:Bands} describes an interface between chiral magnets that prefer skyrmions on the left of the interface and antiskyrmions on the right of the interface.
Here, the instability again can result in the formation of skyrmion-antiskyrmion pair (see Fig.~\ref{creation}).
\begin{figure}[!ht]
\centering
\includegraphics[scale=0.9]{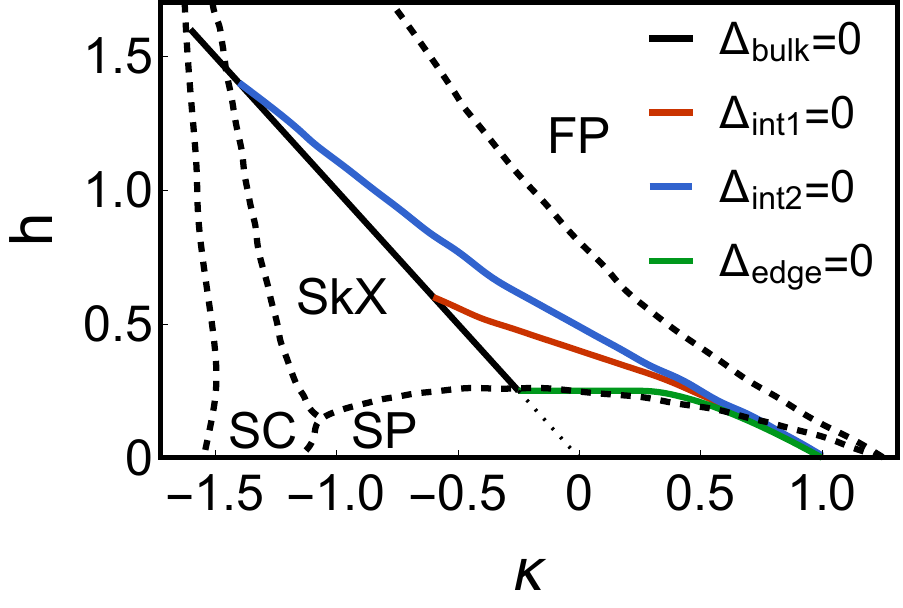}
\caption{(Color online) The phase diagram identifying local instabilities associated with closing of the magnon gap. The dashed lines are reproduced from Refs.~\cite{Utkan,Lin.Saxena.eaPRB2015} and  represent thermodynamic phase boundaries between the polarized state (FP), the hexagonal skyrmion lattice (SkX), the square skyrmion lattice (SC), and the chiral soliton lattice (SP). The black line corresponds to vanishing gap of bulk magnons. The red line corresponds to vanishing gap of magnons localized at an interface between two chiral magnets with $D_3^L/(QJ)=1$, $D_4^L/(QJ)=1$, $D_3^R/(QJ)=1$, $D_4^R/(QJ)=-1$, $\kappa^L=\kappa^R=\kappa$, and $h^L=h^R=h$. The blue line corresponds to vanishing gap of magnons localized at an interface between  a chiral magnet in contact with a non-chiral magnet with $D_3^L/(QJ)=2$, $D_4^L/(QJ)=-2$, $D_3^R/(QJ)=0$, $D_4^R/(QJ)=0$, $\kappa^L=\kappa^R=\kappa$, and $h^L=h^R=h$. The green line corresponds to vanishing gap of magnons localized at an edge of chiral magnet with $D_3/(QJ)=1$, $D_4/(QJ)=0.1$. Dimensionless units are used.}
\label{FigPhaseDiag}
\end{figure}
\subsection{Phase Diagram}
To determine the range of material parameters for which the system can admit non-trivial magnetic structures, it is helpful to draw the stability phase diagram. Such diagram indicate what is likely the state of the system under consideration for the pair of parameters denoted in the abscissa and ordinates. We will be focused mainly on the stability of 3 common magnetic textures: the fully polarized (FP) or ferromagnetic state, the skyrmion (SkX or SC) state and the chiral soliton lattice or spiral (SP) state. The phase diagram for chiral magnets has been drawn in Fig.~\ref{FigPhaseDiag} for the parameters $h$ and $\kappa$ describing the external field and anisotropy respectively. The lines on the diagrams indicates the boundaries of region of similar phase. Such lines, called phase boundaries were determined, in our case, by solving the Bogoliubov-de Gennes equation and ascertaining when the lowest eigenvalue vanish. This closing of the magnon gap indicates a point beyond which the lowest-lying magnon mode can become energetically favorable.

In general, for the formation of topologically non-trivial states it is not sufficient to cross the phase boundaries in Fig.~\ref{FigPhaseDiag} \cite{Togawa.Koyama.eaPRL2012,Wilson.Butenko.eaPRB2014,Keesman.Leonov.eaPRB2015,Utkan}. Instead, an adiabatic change of parameters at low enough temperatures will result in a metastable state. However, the phase boundaries associated with zero gap magnons and local instabilities can lead to formation of chiral solitons which can be broken into skyrmions by magnetic field pulses \cite{Muller-Rosch-Garst,Du.Che.eaNC2015}.
We will now investigate the possible transitions induced by the aforementioned fluctuations. To this end, we study the magnon gap $\Delta$ and identify phase points at which the gap goes to zero.
In particular, we study the magnon gap for bulk magnons $\Delta_{\text{bulk}}$, for interface magnons $\Delta_{\text{int}}$, and for edge magnons $\Delta_{\text{edge}}$ (see Fig.~\ref{FigPhaseDiag}).
For edge magnons with the gap $\Delta_{\text{edge}}$, we consider the case of highly anisotropic DMI (see upper plot in Fig.~\ref{Fig:Bands}). For interface magnons with the gap $\Delta_{\text{int1}}$, we consider the case of a chiral magnet that prefers skyrmions on the left of the interface and antiskyrmions on the right of the interface (see lower plot in Fig.~\ref{Fig:Bands}). We also consider an interface between a magnet with a standard interfacial DMI on one side and no DMI on the other side. The gap of such magnons is denoted by $\Delta_{\text{int2}}$.

The phase diagram in Fig.~\ref{FigPhaseDiag} shows lines at which the magnon gaps vanish, i.e., $\Delta_{\text{bulk}}=0$, $\Delta_{\text{int1}}=0$, $\Delta_{\text{int2}}=0$, and $\Delta_{\text{edge}}=0$. Crossing any of the lines leads to local instability resulting in a non-collinear state. In all four case, we were able to generate skyrmions and/or antiskyrmions with the help of a magnetic pulse as discussed in the next section. 
\begin{figure}[!htb]
\centering
\includegraphics[scale=0.6]{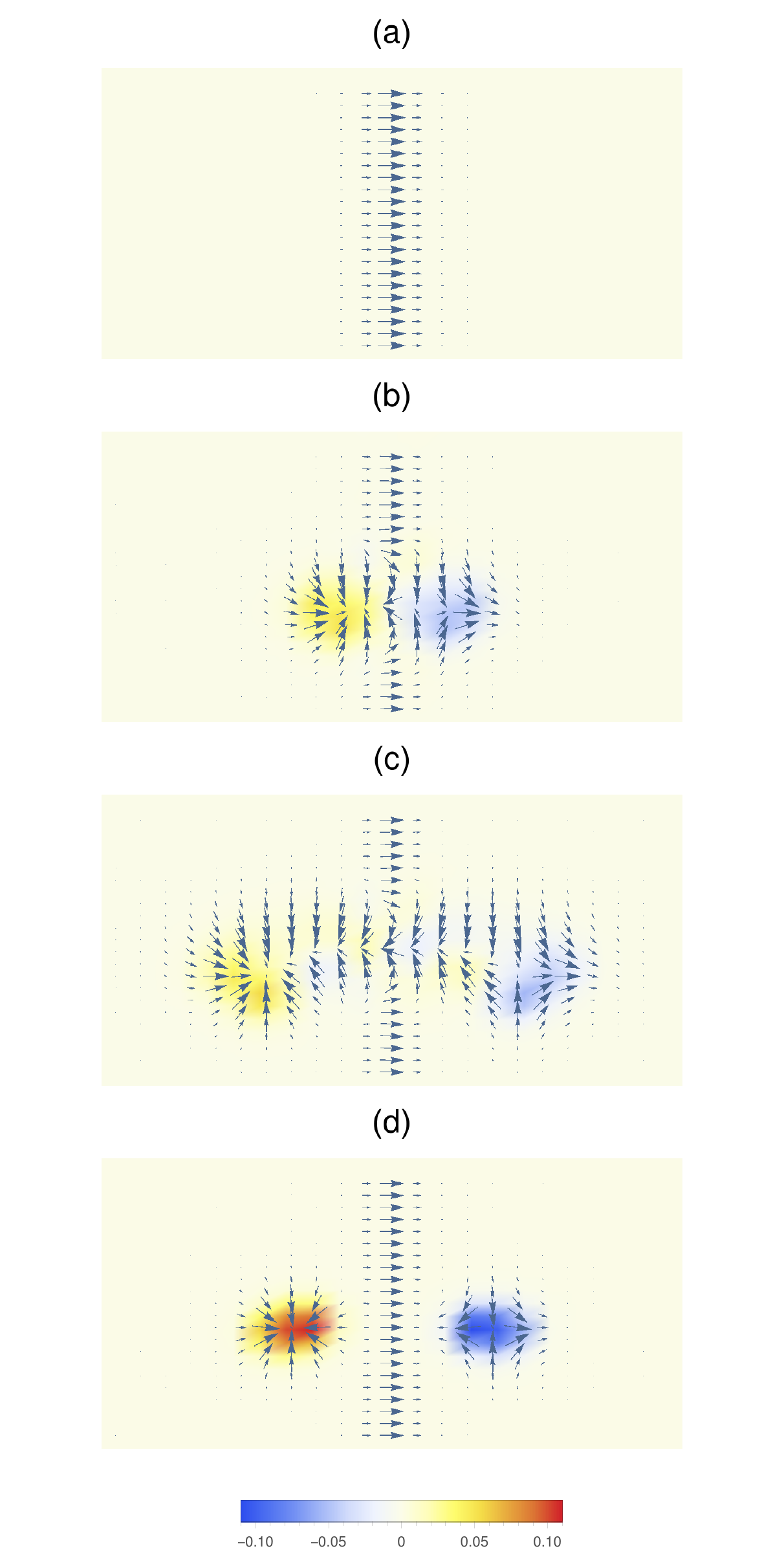}
\caption{(Color online) Snapshots of skyrmion-antiskyrmion pair creation process in a system with an interface separating the Rashba- and Dresselhaus-like DMI which corresponds to the curve $\Delta_{\text{int1}}$ in Fig.~\ref{FigPhaseDiag}. The plots show the in-plane magnetization and the topological charge density.  (a) Initial configuration corresponds to the anisotropy $\kappa= 0$ and magnetic field $h_0=0.9$. (b) and (c) For a period $\Delta t=0.3$~ns the magnetic field pulse lowers the magnetic field to $h_i=0.3$ and leads to formation of bubble-like structure on both sides of the interface. (d) After the magnetic field is returned to initial value the skyrmion and antiskyrmion form from the magnetic bubble.}
\label{creation}
\end{figure}

\begin{figure}[!htb]
\centering
\includegraphics[scale=0.6]{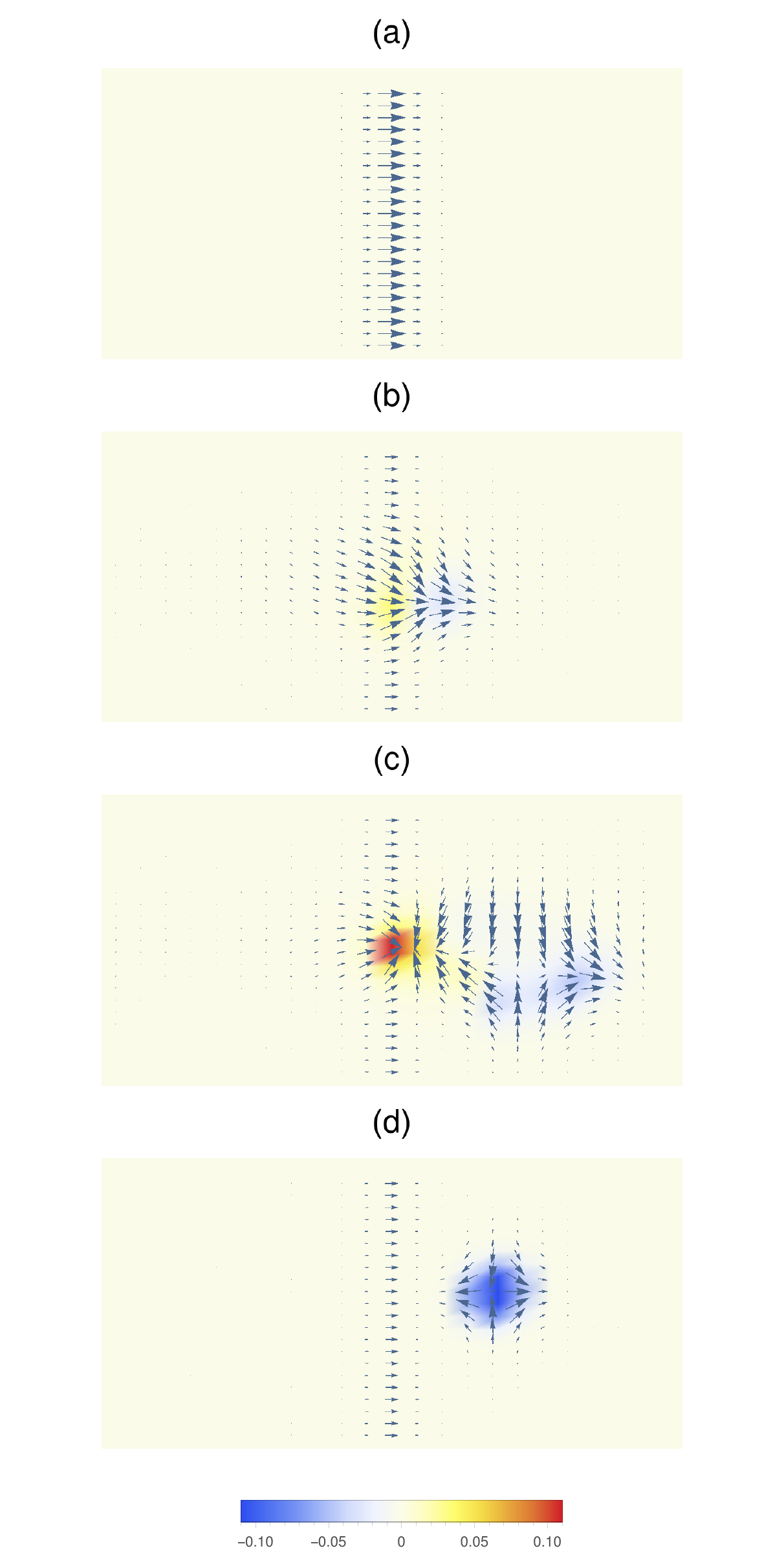}
\caption{(Color online) Snapshots of antiskyrmion creation process in a system with an interface separating a region with no DMI and a region with Dresselhaus-like DMI which corresponds to the curve $\Delta_{\text{int2}}$ in Fig.~\ref{FigPhaseDiag}. The plots show the in-plane magnetization and the topological charge density.  (a) Initial configuration corresponds to the anisotropy $\kappa= 0$ and magnetic field $h_0=0.9$. (b) and (c) For a period $\Delta t=0.3$~ns the magnetic field pulse lowers the magnetic field to $h_i=0.1$ and leads to formation of bubble-like structure on the right-hand side of the interface. (d) After the magnetic field is returned to initial value the antiskyrmion forms from the magnetic bubble.}
\label{creation1}
\end{figure}
\section{Micromagnetics simulations}\label{sec:microM}
To confirm that we can indeed create skyrmions and antiskyrmions using the interface twists, we performed micromagnetics simulations of a thin ferromagnetic film using modified open source micromagnetics simulator mumax3 \cite{Vansteenkiste2014} as well as built-in MATHEMATICA function NDSolve \cite{mathematica}. 

In the first micromagnetics simulation, we consider a ferromagnetic thin film sufficiently long along the $x$-direction ($\sim 1000 \text{nm}$) and sufficiently wide in the $y$-direction ($\sim 200 \text{nm}$), with a thickness of $1\text{nm}$. An interface separates the thin film into two regions with different DMI, namely DMI with $D_{2d}$ symmetry on the right and DMI with $C_{\infty v}$ on the left (this case corresponds to the curve $\Delta_{\text{int1}}$ in Fig.~\ref{FigPhaseDiag}).
In the second micromagnetics simulation, we consider a ferromagnetic thin film with the same geometry in which an interface separates a region with no DMI and a region with $D_{2d}$ symmetry (this case corresponds to the curve $\Delta_{\text{int2}}$ in Fig.~\ref{FigPhaseDiag}).
The ferromagnetic film corresponding to Fig.~\ref{creation} has the following materials parameters: the exchange coupling constant $J/2 = 16~\text{pJ}/m$, the saturation magnetization  M$_{s}$=$10^6$~Am$^{-1}$, the strength of DMI $D_3^L =D_4^L=D_3^R=-D_4^R=4$~mJ/m$^{2}$, and the Gilbert damping $\alpha = 0.3$. The ferromagnetic film corresponding to Fig.~\ref{creation1} is described by the same parameters and by DMI $D_3^L =D_4^L=0$ and $D_3^R=-D_4^R=4$~mJ/m$^{2}$. For simplicity, the presented results were calculated in the absence of any uniaxial anisotropy. Adding uniaxial anisotropy and dipole-dipole interactions does not modify the dynamics substantially.  

In Figs.~\ref{creation} and \ref{creation1}, we plot snapshots of skyrmion and antiskyrmion creation process. To gain additional insight we also plot the topological charge density:
\begin{align}
\rho = \frac{1}{4 \pi} (\partial_x \boldsymbol m \times \partial_y \boldsymbol m)\cdot \boldsymbol m .
\end{align}
To create and stabilize isolated skyrmions and antiskyrmions, we follow the following protocol. First, the system is initialized with spin-polarized ferromagnetic phase on both sides of the interface and relaxed in the presence of magnetic field $h_0$. 
Next, the magnetic field is reduced to the value $h_i$ in a strip of width $60$~nm along the $x$-direction for a period of $\Delta t=0.3$~ns. Note that the finite width of the strip is necessary to break the translational invariance along the $y$-direction.
Once the field is reduced to the value below the gap closing line in Fig.~\ref{FigPhaseDiag}, local instabilities along the interface are created on both sides. The instabilities quickly turn into the helical state creating the bubble-like structures as shown in Figs.~\ref{creation}c) and \ref{creation1}c).  Increase in magnetic field to initial value $h_0$ leads to the detachment of instabilities from the interface. These detached helical instabilities ultimately stabilize as an antiskyrmion or a skyrmion on the left or right side of the interface, respectively, see Figs.~\ref{creation}d) and \ref{creation1}d). Note that in the absence of DMI the local instability does not develop into a helical state as can be seen in the left plot in Fig.~\ref{creation1}c).
We have also demonstrated that this approach works in many other setups as long a boundary or an interface twist is present.

\section{Summary}\label{sec:Summary}
We have demonstrated, both analytically and numerically, the possibility of creating skyrmions and antiskyrmions in chiral magnets with magnetization twists. To properly describe such magnetization twists at interfaces between different magnets (either magnets with different DMI on each side or finite DMI on one side and no DMI on the other side), we have derived the general interface boundary conditions.  Previous studies of systems with non-uniform DMI have not accounted for the full structure of DMI tensor \cite{Diaz-Troncoso,Mulkers.VanWaeyenberge.eaPRB2017}. We have made manifest the crucial role played by the edge or interface and demonstrated that the fluctuations around the equilibrium magnetization, bound to the edge or interface, can lead to local instabilities. It is such local instabilities that lead to creation of skyrmions and antiskyrmions.
Using micromagnetics simulations, we have confirmed our theoretical predictions.
Our results pave the way for further studies of DMI engineering and new phenomena associated with it.
The realization of real systems with symmetries involved in the earlier discussion is not far-fetched given the fact that $C_{2v}$ is the symmetries of (110) bcc or fcc surfaces, and systems like Fe/W(110) \cite{Hoffmann} or Au/Co/W(110)\cite{Camosi.Rohart.ea:PRB2017} has been synthesized and predicted to host the aforementioned topological structures. Additionally, by capping the films with non-magnetic layer \cite{Balk.Kim.eaPrl2017,Wells.Shepley.eaPRB2017} a different symmetry of the DMI can be imposed, hence an interface can be created between two chiral magnets with different DMI but similar exchange. Thin films with a relevant symmetry can also be realized for instance in a (111)-grown thin film of perovskite oxides in the rhombohedral phase, where the transition metals are hexagonally coordinated (e.g. BiFeO$_3$ \cite{Dong2009, Jeong2014}).

\begin{acknowledgments}

We gratefully acknowledge useful discussions with K.~Belashchenko. This work was supported by the U.S. Department of Energy, Office of Science, Basic Energy Sciences, under Award No. DE-SC0014189. 
The computations were performed utilizing the Holland Computing Center of the University of Nebraska.
\end{acknowledgments}

\bibliographystyle{apsrev4-1}
\bibliography{ISkC}

\end{document}